# Phase separation near half-filling point in superconducting compounds

K. Kucab, G. Górski, and J. Mizia

Faculty of Mathematics and Natural Sciences, University of Rzeszów, 35-959 Rzeszów, Poland

**Abstract**

We present the model of superconducting ceramics using the single band extended Hubbard Hamiltonian. We investigate the simultaneous presence of antiferromagnetism (AF) and *d*-wave superconductivity (SC) in the coherent potential (CP) approximation applied to the on-site Coulomb repulsion *U*. We consider the hopping interaction, $\Delta t$, the inter-site charge-charge interaction, *V*, (creating SC), and the single site Hund's type exchange interaction, $F_{in}$, (creating AF). The influence of these interactions on the separation of superconducting and antiferromagnetic phases near the half-filling point is investigated. Results are compared with the experimental data for YBaCuO and NdCeCuO compounds.

## 1. Introduction

The phenomenon of coexistence between AF and SC is actually barely understood despite many efforts. Such compounds as layered superconducting ceramics (e.g. YBCO, NdCeCuO), Fe-based oxygen pnictides [1], some kinds of heavy-fermion systems [2] and organic superconductors [3], exhibit the coexistence between these two phases. Additionally, carrier doping and external pressure has great influence on the transition between the AF and SC phases. The experimental data shows that at half-filling most of the HTSC's are antiferromagnetic insulators. Doping causes vanishing of the AF state and appearance of SC in the metallic state.

A great amount of theoretical work has been done to describe the phenomenon of competition between AF and SC. The best known and often used methods are: the mean-field approximation ([4-6]), dynamical mean-field theory (DMFT) [7], cluster DMFT [8], the cellular dynamical mean-field theory (CDMFT) [9], the dynamical cluster approximation (DCA) [10] and the spectral density approach (SDA) [11].

In our work we try to reproduce the experimental data using the itinerant extended Hubbard model in the mean-field and CP approximation.

## 2. Model Hamiltonian

We use the single band Hamiltonian. It includes the on-site Coulomb repulsion $U$, the inter-site charge-charge interaction $V$ [12-14], which is responsible for creating *d*-wave SC, the single site Hund's type exchange interaction $F_{in}$, which is responsible for creating AF and the hopping interaction $\Delta t$ (see [15-17]). The Hamiltonian can be written as

$$H = -\sum_{<ij>\sigma}\left[t - \Delta t\left(\hat{n}_{i-\sigma} + \hat{n}_{j-\sigma}\right)\right]c^+_{i\sigma}c_{j\sigma} - \mu_0 \sum_{i\sigma}\hat{n}_{i\sigma}$$
$$+ \frac{U}{2}\sum_{i\sigma}\hat{n}_{i\sigma}\hat{n}_{i-\sigma} + \frac{V}{2}\sum_{<ij>}\hat{n}_i\hat{n}_j - F_{in}\sum_{i\sigma}n_\sigma\hat{n}_{i\sigma} ,$$

(1)

where $\mu_0$ is the chemical potential, $c_{i\sigma}^+ (c_{i\sigma})$ are the electron creation (annihilation) operators, $\hat{n}_{i\sigma} = c_{i\sigma}^+ c_{i\sigma}$ is the electron number operator, $\hat{n}_i = \hat{n}_{i\sigma} + \hat{n}_{i-\sigma}$, and $n_\sigma$ is the probability of finding the electron with spin $\sigma$ in a band. Parameter $t$ is the hopping amplitude, and $\Delta t$ is the hopping interaction. The interaction $F_{in}$ lowers the energy of each pair of electrons with parallel spins with respect to antiparallel pairs. The term proportional to $V \sum_{<ij>} \hat{n}_i \hat{n}_j$ (with negative $V$) comes from the reduction of the three-band extended Hubbard model to the effective single-band model. The existing $d$-wave superconductivity (see [12-14]) is created by the negative inter-site charge-charge interaction $V$, and it is not hindered by the repulsion $U$. The concentration dependence of the effective bandwidth, $D(n)$, is realized due to the nonzero value of $\Delta t$ interaction.

Further analysis is based on equations of motion for the Green functions. We divide the crystal lattice into two interpenetrating magnetic sub-lattices $\alpha, \beta$. Next we use the Hartree-Fock approximation to the inter-site interactions $\Delta t$ and $V$, and the CP approximation to $U$. We consider $d$-wave SC, the AF ordering and their mutual interaction.

## 3. Numerical results

Firstly, we investigate the influence of interaction $\Delta t$ on the range of the existence of the AF state.

The curve in Fig. 1 presents values of carrier concentration at which the AF state disappears for different values of $\Delta t$ interaction. The value of $F_{in}$ is fitted to the Néel's temperature of 500K at $n = 1$. This figure allows us to investigate the influence of the $\Delta t$ interaction on the range of existence of antiferromagnetism. In the current work we assume the initial half-bandwidth of the unperturbed DOS (for 2D square lattice) equal to $D_0 = 2\text{eV}$.

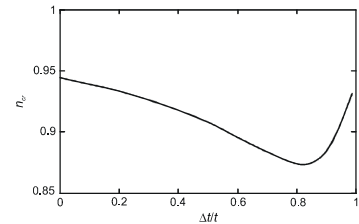

Fig. 1 The critical values of AF carrier concentration $n_{cr}$ (where the AF state disappears) as a function of $\Delta t$ interaction (full description in the text). The other parameters used in computations are: $U = 0.17 eV$, $V = -0.18\text{eV}$.

As one can see with the increase of interaction $\Delta t$ the AF state persists from $n = 1$ to the smaller concentrations region, which contradicts the experimental data, according to which, the AF state exists only very close to the half-filling point. At relatively high values of the $\Delta t$ interaction ($\sim 0.8t$) the region of AF state starts to decrease towards the half-filled point. Unfortunately, even a very high value of $\Delta t$ interaction ($\sim 0.99t$) cannot improve the situation as compared to the system without hopping interaction, where the range of occurrence of the AF state about the half-filled point is 27% smaller.

Taking into account the significant influence of $\Delta t$ on the SC state and affecting the SC state positively by pushing it away from the half filled point (see Figs 2 and 3 below) we will use the $\Delta t$ interaction (accepting its negative influence on AF state) in further calculations.

In Fig. 2 we present superconducting critical temperatures $T_{cr}^{SC}$ for the d-wave SC and compare this with the experimental data for the YBCO compound in the function of carrier concentration $n$ for different values of interaction $V$. The black solid lines are calculated for the same parameters as the red dashed lines in Fig. 3 below.

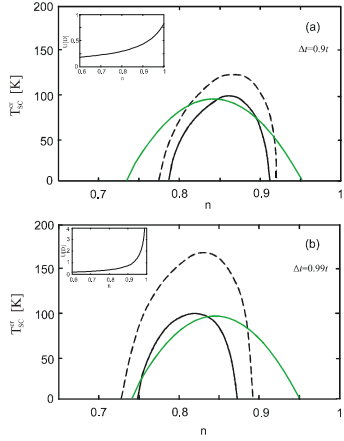

Fig. 2 Comparison with experimental data (green curves) of SC critical temperatures in function of carrier concentration. The interactions are: (a) $F_{in} = 0.15\,\text{eV}$, $V = -0.18\,\text{eV}$ (solid line), $V = -0.184\,\text{eV}$ (dashed line); (b) $F_{in} = 0.1\,\text{eV}$, $V = -0.176\,\text{eV}$ (solid line), $V = -0.188\,\text{eV}$ (dashed line). For both graphs $U = 0.17\,\text{eV}$. Inset: the effective Coulomb interaction as a function of carrier concentration.

As can be seen from above figures, the size of the superconducting effect is very sensitive to the strength of the negative charge-charge interaction. The SC critical curve is pushed away from the half filling point, which reflects the realistic experimental data e.g. for the YBCO compound.

In Fig. 3 we show the results for SC and AF phases considering their mutual interaction. The numerical results (red curves) are compared with the experimental data (green curves).

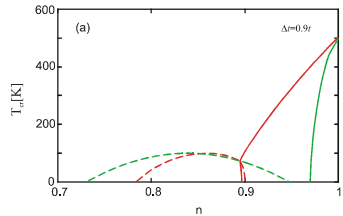
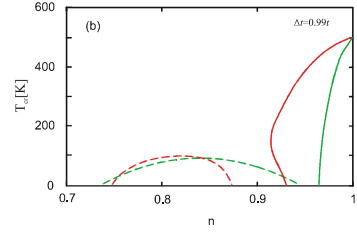

Fig. 3 Comparison of numerical results and experimental data (green curves) for the YBa$_2$Cu$_3$O$_7$ compound. The values of interactions used in the computations are: (a) $F_{in} = 0.15\,\text{eV}$, $V = -0.18\,\text{eV}$; (b) $F_{in} = 0.1\,\text{eV}$, $V = -0.176\,\text{eV}$. For both graphs $U = 0.17\,\text{eV}$.

As we can see, the theoretical curves for AF ordering still depart from the experimental data. This could be an effect because we used an insufficient approximation (CPA), which in the vicinity of the half-filled point and at relatively strong $U/D(n)$ ratio does not satisfactorily describe the AF correlation.

As we know, the hopping interaction, $\Delta t$, is responsible for breaking the electron-hole symmetry. Taking into account this asymmetry we show in Fig. 4 the critical temperatures $T_{SC}$ and $T_N$ for the hole-doped ($n<1$) and also for the electron-doped ($n>1$) region.

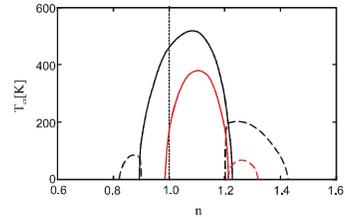

Fig. 4 Neel's temperatures (solid lines) and SC critical temperatures (dashed lines) in the function of carrier concentration for $\Delta t = 0.4t$ and $U = 0.4\,\text{eV}$. The remaining parameters are: $V = -0.42\,\text{eV}$, $F_{in} = 0.43\,\text{eV}$ (black curves), and $V = -0.4\,\text{eV}$, $F_{in} = 0.41\,\text{eV}$ (red curves).

Fig. 4 shows that the stimulation of AF and SC by the hopping interaction is stronger for electron-doped ($n>1$) compounds than for hole-doped ones. At slightly smaller values of charge-charge interactions (the red curves) it is even possible for a given compound not to have a SC state at the hole-doped concentrations ($n<1$), and to obtain a good fit to the experimental results at $n>1$. This would be possibly the case of Nd$_{2-x}$Ce$_x$CuO$_{4-y}$ compound.

## 4. Conclusions

The results we obtained are quite close to the experimental data. The advantage of our approach is, that by using a relatively simple CP approximation (with respect to very tedious numerical methods e.g. Quantum Monte-Carlo DMFT), we obtain the phase diagrams closer to the experimental data than the methods mentioned.

As we can see, the mutual damping of antiferromagnetic and superconducting states is strong. The region of coexistence between these states (if the coexistence appears) is very narrow. When we increase the $\Delta t$ interaction to relatively high values, the effective Coulomb interaction represented by the $U/D(n)$ ratio increases (due to decrease of the effective bandwidth) and the SC state is repelled towards smaller concentrations, due to the large shape modification of the initial DOS. The increase of the effective Coulomb interaction can even separate the appearance in the concentration range of the AF and SC phases.

Taking into account the *d*-wave superconductivity we can suitably adjust the phase diagram to the Neel's temperature and to the experimental data for the YBCO compound.

The influence of the hopping interaction on the AF and SC orderings is much stronger in the electron-doped region ($n>1$) than in the hole-doped concentration range ($n<1$). We can choose the set of interactions for which the SC phase will disappear at the hole-doped region and occure at the electron-doped region, which is in agreement with the experimental data, e.g. for the NdCeCuO compound.

## References

[1] J. Dong, H.J. Zhang, G. Xu, Z. Li, G. Li, W.Z. Hu, D. Wu, G.F. Chen, X. Dai, J.L. Luo, Z. Fang and N.L. Wang, Nature 83, (2008) 27006.
[2] C. Pfleiderer, Rev. Mod. Phys. 81, (2009) 1551.
[3] J. Wosnitza, J. Low Temp. Phys. 146, (2007) 641.
[4] K. Machida and M. Ichioka, J. Phys. Soc. Jpn. 68, (1999) 4020.
[5] B. Kyung, Phys. Rev. B 62, (2000) 9083.
[6] J. Mizia, G. Górski and K. Kucab, phys. stat. sol. (b) 229, No. 3, (2002) 1221.
[7] A. Georges, G. Kotliar, W. Krauth and M.J. Rozenberg, Rev. Mod. Phys. 68, (1996) 13.
[8] A. I. Lichtenstein and M. I. Katsnelson, Phys. Rev. B 62, (2000) R9283.
[9] S.S. Kancharla B. Kyung, D. Sénéchal, M. Civelli, M. Capone, G. Kotliar and A.-M.S. Tremblay, Phys. Rev. B 77, (2008) 184516.
[10] T. Maier, M. Jarrell, T. Pruschke and M.H. Hettler, Rev. Mod. Phys. 77, (2005) 1027
[11] T. Herrmann and W. Nolting, J. Magn. Magn. Mater. 170, (1997) 253.
[12] J. Mizia and G. Górski, *Models of itinerant ordering in crystals: An introduction*, Elsevier Ltd., Amsterdam 2007.
[13] T. Domański and K.I. Wysokiński, Phys. Rev. B 59, (1999) 173.
[14] W. Weber, Z. Phys. B 70, (1988) 323.
[15] J.E. Hirsch, F. Marsiglio, Phys. Rev. B 39, (1989) 11515.
[16] F. Marsiglio and J.E. Hirsch, Phys. Rev. B 41, (1990) 6435.
[17] G. Górski and J. Mizia, Phys. Rev. B 83, (2011) 064410